\documentclass[12pt]{article}
\usepackage{amsmath,amsfonts,amssymb}
\usepackage{graphicx}
\usepackage{psfrag}
\usepackage{enumerate}

\numberwithin{equation}{section}

\begin{document}
\newcommand{\todo}[1]{{\em \small {#1}}\marginpar{$\Longleftarrow$}}
\newcommand{\labell}[1]{\label{#1}\qquad_{#1}} 

\rightline{DCPT-07/19}
\vskip 1cm

\begin{center}
{\Large \bf Geometry of non-supersymmetric three-charge bound states}
\end{center}
\vskip 1cm

\renewcommand{\thefootnote}{\fnsymbol{footnote}}
\centerline{\bf Eric G. Gimon$^a$\footnote{eggimon@lbl.gov}, Thomas S. Levi$^b$\footnote{levi@physics.nyu.edu} and Simon
F. Ross$^c$\footnote{S.F.Ross@durham.ac.uk}}
\vskip .5cm
\centerline{\it $^a$Berkeley Center for Theoretical Physics, Berkeley, CA 94720, U.S.A.}
\vskip .5cm
\centerline{\it $^b$Center for Cosmology and Particle Physics}
\centerline{\it New York University, 4
  Washington Place, New York, NY 10003,  U.S.A.} 
\vskip .5cm
\centerline{ \it $^c$Centre for Particle Theory, Department of
Mathematical Sciences}
\centerline{\it University of Durham, South Road, Durham DH1 3LE, U.K.}

\setcounter{footnote}{0}
\renewcommand{\thefootnote}{\arabic{footnote}}

\vskip 1cm
\begin{abstract}
We study the smooth non-supersymmetric three-charge microstates of
Jejjala, Madden, Ross and Titchener~\cite{rossnonsusy} using Kaluza-Klein
reductions of the solutions to five and four dimensions.  Our aim is
to improve our understanding of the relation between these
non-supersymmetric solutions and the well-studied supersymmetric
cases. We find some surprising qualitative differences. In the
five-dimensional description, the solution has orbifold fixed points
which break supersymmetry locally, so the geometries cannot be
thought of as made up of separate half-BPS centers. In the four-dimensional
description, the two singularities in the geometry are connected by
a conical singularity, which makes it impossible to treat them
independently and assign unambiguous brane charges to these
centers.

\end{abstract}

\section{Introduction}

The study of black holes and their thermodynamics has led to a number
of important advances in string theory. Recently, the construction of
smooth geometries corresponding to individual microstates of the black
holes (more precisely, in a dual field theory description, to
individual pure states which contribute to the thermal ensemble) has
received extensive attention. Supersymmetric two-charge microstates
are now reasonably
well-understood~\cite{juanliat,conicaldefect,mathurmultstring,mathurhottube,mathuradsparadox,LLM,skend1,skend2,skend3}
(see~\cite{mathur1,mathur2} for a review), and substantial progress
has been made on understanding the supersymmetric three-charge
microstates~\cite{mathurhair,mathura,mathurb,mathur3,us,bw2,benawarnerfoaming,Ford:2006yb,Bena:2006kb} and the dimensional
reduction to four dimensions~\cite{per2,per3,4dmicro,Saxena:2005uk}. Our aim in this paper is
to continue the investigation of non-supersymmetric states initiated
in~\cite{rossnonsusy}.

In the three-charge case, an important step was the
rewriting~\cite{mathur3} of the first examples of smooth three-charge
geometries in the fibered form used in the classification of
supersymmetric solutions~\cite{reall5d}.  This led to the realisation that the base
space for these solutions had a ``pseudo-hyper-K\"ahler'' form, which
led to the generalisations to multi-center solutions~\cite{us,bw2}.
These solutions all have a $U(1)$ isometry in the base. This may
appear to be an unnatural restriction for the five-dimensional case,
but if we add a NUT charge and reduce along the isometry direction,
this family of solutions naturally corresponds to microstates of
four-dimensional asymptotically flat black
holes~\cite{per2,per3,4dmicro}. This focus on the centers was refined
in~\cite{4dmicro}, where it was conjectured that the general
four-dimensional supersymmetric microstate is made up of half-BPS
``atoms''. This will be described by a multi-center geometry if the
distance between the individual ``atoms'' is sufficiently large, or by
passing to an open string description if the separations are small. This separation is modulated by the string coupling.

We would like to apply the lessons learnt from this analysis to the
non-supersym\-metric solutions of~\cite{rossnonsusy}. These solutions
are qualitatively similar to the simplest supersymmetric two-center
three-charge solutions studied in~\cite{mathura,mathurb,mathur3}, so
analysing them in a similar way should shed some light on the
similarities and differences, and ultimately guide us towards
constructing more general non-supersymmetric microstates. The plan is
thus to first Kaluza-Klein reduce the solutions to five dimensions,
where we can see that the solutions have two centers. We will give a
description of the five-dimensional solution as a $T^6$
compactification from M-theory, which makes the symmetry between the
charges manifest. This five-dimensional solution has a $U(1)$ isometry
in the base. It is natural to consider a further dimensional
reduction to four dimensions along this isometry to attempt to obtain
a more direct physical interpretation, as
in~\cite{4dmicro}.

Implementing this program leads to some surprises: first, in the
five-dimensional solution, the two centers are locally described by
non-supersymmetric orbifolds. We had expected to find an orbifold
singularity at these points in the five dimensional description, but
it is a surprise that the supersymmetry is broken even locally: in the
six-dimensional description, the full asymptotically flat solution is
non-supersymmetric, but it has (for suitable choices of parameters) a
near-core AdS$_3 \times S^3$ geometry, in which supersymmetry is
restored. This supersymmetry is broken by the choice of Kaluza-Klein
reduction: even if we consider the Kaluza-Klein reduction from the
exact AdS$_3 \times S^3$ geometry, the five-dimensional solution
contains non-supersymmetric orbifold singularities. This leads to the
first general lesson we wish to draw: the picture of microstates as
made up of half-BPS ``atoms'' does {\it not} extend to the
non-supersymmetric case. We must consider more general kinds of basic
building blocks.

If we reduce to four dimensions, we find our second surprise: the
four-dimensional solution is not smooth away from the centers. There
is a conical singularity which connects the two centers. The ambiguity
in the definition of the Kaluza-Klein gauge field on this conical line
singularity makes it impossible for us to unambiguously associate
brane charges with the two centers.  This suggests that unlike in the
supersymmetric case, we cannot treat the two centers independently,
and the physical interpretation of these centers in the
four-dimensional description is much less clear here.

One of the ultimate goals of this work was to make progress towards
constructing more general non-supersymmetric solutions. In general,
constructing non-supersym\-metric solutions is much more difficult than
supersymmetric ones. In the non-supersymmetric case we have to deal
directly with the non-linear, second order equations of motion, which
makes it impossible to construct solutions in terms of harmonic
functions, as can be done in the supersymmetric case. The results we
have obtained show that the non-supersymmetric case is quite different
from the supersymmetric one, and suggest that it will be difficult to
extend the work on supersymmetric cases to construct multi-center
non-supersymmetric solutions. However, it may still be possible at
least to construct non-supersymmetric solutions which are
asymptotically flat in four dimensions; work on this problem is
ongoing~\cite{grs}.

The plan of the rest of the paper is as follows: in the next section,
we review the non-supersymmetric solitons of~\cite{rossnonsusy} in
type IIB supergravity. In section~\ref{mth}, we convert these
solutions to an M-theory form, write the metric in terms of the
integer parameters of the solitons, and discuss the relation to the
supersymmetric case. In section~\ref{str}, we analyse the structure of
the M-theory solutions, with a particular focus on the orbifold
singularities appearing in this five-dimensional description. Finally,
in section~\ref{4d}, we discuss the dimensional reduction to four
dimensions.

\section{Review of construction}

We start from the metric for the D1-D5-P black string, written in the
fibered form in~\cite{rossnonsusy},
\begin{align} \label{fibred}
\mathrm{d}s^2 &= \frac{1}{\sqrt{ \tilde{H_1}\tilde{H_5} } }\left\{
-(f-M) \left[\mathrm{d}\tilde{t}- (f-M)^{-1} M
\cosh\delta_1\cosh\delta_5 (a_1 \cos^2\bar \theta \mathrm{d}\psi + a_2
\sin^2 \bar \theta 
\mathrm{d}\phi) \right]^2 \right. \nonumber \\ &+ f \left.\left[
\mathrm{d}\tilde{y}+f^{-1} M \sinh\delta_1\sinh\delta_5 (a_2
\cos^2\bar \theta \mathrm{d}\psi + a_1 \sin^2\bar \theta \mathrm{d}\phi) \right]^2
\right\}\nonumber\\ &+ \sqrt{\tilde{H_1}\tilde{H_5}}\left\{
\frac{\bar{r}^2\mathrm{d}\bar{r}^2}{
  (\bar{r}^2+a_1^2)(\bar{r}^2+a_2^2)-M\bar{r}^2 } 
+\mathrm{d}\bar \theta^2 \right.\nonumber \\ &+ (f(f-M))^{-1}\left[
  \left(f(f-M)+f a_2^2\sin^2\bar \theta - (f-M)a_1^2\sin^2\bar \theta
  \right)\sin^2\bar \theta \mathrm{d}\phi^2 \right. \nonumber \\ &+ 2 M
  a_1 a_2 \sin^2 \bar \theta \cos^2 \bar \theta \mathrm{d}\psi \mathrm{d}\phi
  \nonumber \\
  &+ \left.\left.\left(f(f-M)+fa_1^2\cos^2\bar \theta -
  (f-M)a_2^2\cos^2\bar \theta\right)\cos^2\bar \theta \mathrm{d}\psi^2
  \right] \phantom{ \frac{1}{1}} \right\} + \sqrt{\frac{\tilde
    H_1}{\tilde H_5}} \sum_{i=1}^{4} dz_i^2,
\end{align}
where $\tilde t = t \cosh \delta_p - y \sinh \delta_p$, $\tilde y = y
\cosh \delta_p -t \sinh \delta_p $, 
\begin{equation} 
\tilde{H}_{i}=f+M\sinh^2\delta_i, \quad
f=\bar r^2+a_1^2\sin^2 \bar \theta+a_2^2\cos^2\bar \theta.
\end{equation}
This metric is more usually written in terms of functions $H_i =
\tilde H_i/f$.  Writing it in this way instead makes it clear that
there is no singularity at $f=0$. The above metric is in the string
frame, and the dilaton is
\begin{equation}
e^{2\Phi} = \frac{\tilde H_1}{\tilde H_5}.
\end{equation}
{}The 2-form gauge potential which supports this
configuration is
\begin{align} \label{rr}
C_2 &= \frac{M \cos^2 \bar \theta}{\tilde H_1} \left[ (a_2 c_1
  s_5 c_p - a_1 s_1 c_5
  s_p) dt + (a_1 s_1 c_5 c_p - a_2 c_1 s_5 s_p) dy \right] \wedge d\psi
   \\
&+ \frac{M \sin^2 \bar \theta}{\tilde H_1} \left[  (a_1 c_1
  s_5 c_p - a_2 s_1 c_5
  s_p) dt  + (a_2 s_1 c_5 c_p - a_1 c_1 s_5 s_p) dy \right] \wedge d \phi
  \nonumber \\
&- \frac{M s_1 c_1}{\tilde H_1} dt \wedge dy -
  \frac{M s_5 c_5}{\tilde H_1} (\bar r^2 + a_2^2 + M
  s_1^2) \cos^2 \bar \theta d\psi \wedge d\phi, \nonumber
\end{align}
where $c_i = \cosh \delta_i$, $s_i = \sinh \delta_i$. We take the
$T^4$ in the $z_i$ directions to have volume $V$, and the $y$ circle
to have radius $R$, that is $y \sim y + 2\pi R$.

This metric and fields describes a family of solutions with D1, D5 and
P charges, labelled by the seven parameters $M, a_1, a_2, \delta_1,
\delta_2, \delta_3, R$. It was shown in~\cite{rossnonsusy} that the
geometry is smooth everywhere and contains no closed timelike curves
if these parameters are fixed in terms of two integers $m,n$ and the
charges $q_1, q_5, q_p$ by the relations
\begin{equation} \label{mass}
M = a_1^2+a_2^2-a_1 a_2 \frac{(c_1^2 c_5^2 c_p^2 + s_1^2 s_5^2 s_p^2
  )}{s_1 c_1 s_5 c_5 s_p c_p},
\end{equation}
\begin{equation}
\frac{s_p c_p }
{(a_1 c_1 c_5 c_p- a_2 s_1 s_5 s_p)} R = n, \quad \frac{s_p c_p } 
{(a_2 c_1 c_5 c_p- a_1 s_1 s_5 s_p)} R = -m,
\end{equation}
\begin{equation} \label{period}
R =\frac{ M s_1 c_1 s_5 c_5 (s_1 c_1 s_5 c_5 s_p
  c_p)^{1/2}}{\sqrt{a_1a_2}(c_1^2 c_5^2 c_p^2 - s_1^2 s_5^2 s_p^2 )}.
\end{equation}
If we introduce dimensionless parameters 
\begin{equation} \label{params}
j = \left( \frac{a_2}{a_1} \right)^{1/2} \leq 1 , \quad s = \left( \frac{ s_1
  s_5 s_p}{ c_1 c_5 c_p} \right)^{1/2} \leq 1, 
\end{equation}
then the integer quantisation conditions determine these via
\begin{equation} \label{integer}
\frac{ j+ j^{-1}}{s + s^{-1}} = m-n, \quad \frac{ j - j^{-1}}{s -
  s^{-1}} = m+n, 
\end{equation}
Which gives
\begin{equation} \label{smn}
s^2 = \frac{m^2+n^2-1 - \sqrt{(m^2-n^2)^2-2(m^2+n^2)+1}}{2mn}.
\end{equation}
Note that this constraint is invariant under the permutation of the
three charges. We can rewrite the mass \eqref{mass} as
\begin{equation}
M = a_1 a_2 (s^2 -j^2)(j^{-2} s^{-2} - 1) = a_1 a_2 nm (s^{-2} - s^2)^2,
\end{equation}
so having solved for $s$ in terms of $m, n$, we can use this to write
$a_1 a_2$ in terms of $M$. The only remaining step is to solve for $M$
in terms of $m, n$ and the charges. This was left entirely implicit
in~\cite{rossnonsusy}; we will say a little more about it below.

An unsatisfactory feature of this story is the highly implicit nature
of the above conditions. We will see below that we can write the
metric explicitly in terms of the integer quantised parameters $m,n$
and a single length scale, which is related to $M$ and hence
determined in terms of the charges $q_i$. This will bring the analysis
much closer to the supersymmetric cases. It is worth pointing out at
this stage that it would equally have been possible to rewrite it in
this way in the type IIB form above; this rewriting is independent of
the transformation to the M-theory form.

In what follows, we will usually assume $m,n$ are coprime. If
we do not, there will be additional orbifold singularities in the
M-theory form of the solution. 

\section{M-theory form}
\label{mth}

The general study of solutions in the supersymmetric case is conducted
in an M-theory picture, where the symmetry between the three charges
is manifest. Our first task is therefore to translate the type IIB form in
which the non-supersymmetric solutions were first obtained to an M-theory form.

To pass to the M-theory form, we T-dualise on $z_3, z_4$ to get to a
D3-D3-P solution, and then T-dualise on $y$ to get to D2-D2-F1 in IIA.
The IIA solution is then uplifted to M-theory. There is a
symmetry between the charges in the M-theory picture, so we rename the
charges to $Q_1 = q_1, Q_2 = q_5 , Q_3 = q_p$. For the general family
of rotating black string solutions~\eqref{fibred}, the resulting
solution is
\begin{equation} 
ds^2_{11} = - (\tilde H_1 \tilde H_2 \tilde H_3)^{-2/3} f (f-M) (dt
+ k)^2 + (\tilde H_1 \tilde H_2 \tilde H_3)^{1/3} d\bar{s}_4^2 + ds^2_{T^6},
\end{equation}
where
\begin{equation}
k = - \frac{M c_1 c_2 c_3}{f-M} (a_1 \cos^2 \bar \theta d\psi + a_2 \sin^2
\bar \theta d \phi) + \frac{M s_1 s_2 s_3}{f} (a_2 \cos^2 \bar \theta d\psi +
a_1 \sin^2 \bar \theta d\phi),
\end{equation}
and the four-dimensional base metric is 
\begin{align} \label{4dbasegen}
\bar{ds}_4^2 &= \frac{\bar r^2 d\bar r^2}{g(\bar r)} + d\bar \theta^2 \\
&+ (f(f-M))^{-1} [ (f(f-M) + f(a_2^2-a_1^2) \sin^2 \bar \theta + M a_1^2
  \sin^2 \bar \theta) \sin^2 \bar \theta d\phi^2 \nonumber \\ 
&+ 2M a_1 a_2 \sin^2 \bar \theta \cos^2 \bar \theta d \phi d \psi \nonumber \\
&+ ( f(f-M) + f(a_1^2-a_2^2) \cos^2 \bar \theta + M a_2^2 \cos^2 \bar
\theta)
  \cos^2 \bar \theta d\psi^2 ], \nonumber
\end{align}
where
\begin{equation}
  g(\bar r) = (\bar r^2+a_1^2)(\bar r^2+a_2^2)-M\bar r^2 \equiv (\bar
  r^2 - r_+^2)(\bar r^2 - r_-^2), \quad f = \bar r^2 + a_1^2 \sin^2
  \bar \theta + a_2^2 \cos^2 \bar \theta. 
\end{equation}
Note that the coordinate $\bar r$ lives on the interval $[r_+,\infty)$; this fact will become clearer in some of the later coordinate systems. The metric is asymptotically flat in five dimensions. The coordinates
$\bar \theta, \phi, \psi$ are conventional coordinates on the $S^3$ at
large $\bar r$. The metric on the $T^6$, which plays little part in
our discussions, is
\begin{equation}
  ds^2_{T^6} = (\tilde H_1 \tilde H_2 \tilde H_3)^{1/3} (\tilde
  H_1^{-1} (dz_1^2 + dz_2^2) + 
  \tilde H_2^{-1} (dz_3^2 + dz_4^2) + \tilde H_3^{-1} (dz_5^2 + dz_6^2)).
\end{equation}
The gauge field in eleven dimensions is
\begin{equation}
C_{(3)} = A_1 \wedge dz_1 \wedge dz_2 + A_2 \wedge dz_3 \wedge dz_4 +
A_3 \wedge dz_5 \wedge dz_6, 
\end{equation}
where
\begin{equation}
  A_i = -\frac{1}{\tilde H_i} [Q_i dt + M c_j c_k s_i (a_1 \cos^2
  \bar \theta d\psi + a_2 \sin^2 \bar \theta d \phi) - M s_j s_k c_i (a_2 \cos^2
  \bar \theta d\psi +  a_1 \sin^2\bar \theta d\phi)],  
\end{equation}
where here and in similar formulae throughout, $j$ and $k$ denote the
other two charge parameters, $i\neq j \neq k$ (so if $i=1$, $j=2$ and
$k=3$, and so forth). The charges are defined as 
\begin{equation}
Q_i = M \sinh
\delta_i \cosh \delta_i.
\end{equation}

\subsection{Adapted coordinates}

The foregoing discussion applies to the general family of
solutions~\eqref{fibred}; we would like to specialise to the
smooth soliton solutions, and eliminate the dependence on the
parameters to rewrite this solution in a way which makes the
dependence on the integer parameters $m,n$ manifest. This will be
helpful in understanding the relation to the supersymmetric cases, and
will also make the structure of our non-supersymmetric solution
clearer. In this subsection, we introduce a new coordinate system
adapted to this solution. In the next, we will consider coordinates
which are as close as possible to the coordinates used in the
supersymmetric cases.

We set
\begin{equation}
\rho^2 = \frac{\bar r^2 - r_+^2}{r_+^2 - r_-^2}, \quad x = \cos 2\bar
\theta.
\end{equation}
The range of the coordinates is $\rho \geq 0$, $-1 \leq x \leq 1$. In
these coordinates,
\begin{equation}
\frac{\bar r^2 d\bar r^2}{g(\bar r)} = \frac{d\rho^2}{\rho^2+1}. 
\end{equation}
and
\begin{equation}
  f - \frac{1}{2} M = (r_+^2 - r_-^2) \rho^2+ \frac{1}{2} (a_2^2-a_1^2) x + r_+^2 +
  \frac{1}{2} (a_1^2 + a_2^2 -M) = 2c (2 \rho^2 +1 -(m^2-n^2) x ),
\end{equation}
where
\begin{equation} \label{cdef}
c \equiv \frac{r_+^2 - r_-^2}{4}.
\end{equation}
Thus, 
\begin{equation}
V \equiv \frac{f(f-M)}{4c^2} = 4\rho^2(\rho^2+1) - (m^2-n^2)^2 (1-x^2) - 2(m^2-n^2) x
  (2\rho^2+1) + 2 (m^2+n^2). 
\end{equation}

This now involves only the integer parameters $m,n$ and the single length
scale $c$. We can continue to calculate the full metric, and we find
that we can in a similar fashion eliminate all the dependence on the
parameters in favour of $m,n$ and the single length scale $c$ by fully
exploiting (\ref{mass}-\ref{period}). Some useful relations which
follow from (\ref{mass}-\ref{period}) are
\begin{equation}
M a_1 a_2 = 16 c^2 nm, \quad a_1^2 - a_2^2 = 4c (m^2-n^2),
\end{equation}
\begin{equation}
\frac{1}{2} M (a_1^2+a_2^2) = 8c^2 [ (m^2-n^2)^2 - (m^2+n^2)], 
\end{equation}
\begin{equation}
\frac{1}{4} M^2 = 4c^2 [ (m^2-n^2)^2 - 2(m^2+n^2) + 1].
\end{equation}

Omitting the details of the calculation, we find
\begin{align}
  k = \frac{\sqrt{Q_1 Q_2 Q_3}}{4c V \sqrt{mn}} & \left\{ [2\rho^2-(m^2-n^2)(1+x)
   ] n (1-x) d\phi \right. \nonumber \\ &- \left. [2\rho^2+(m^2-n^2)(1-x)] m (1+x) d\psi \right\},
\end{align}
and the base metric is 
\begin{align} \label{xybase}
\bar{ds}_4^2 = & \frac{d\rho^2}{\rho^2+1} + \frac{dx^2}{4(1-x^2)} +
\frac{2nm}{V} (1-x^2) d\phi d\psi \\ &+
\frac{1}{2V} \left[
  4\rho^2(\rho^2+1)-(m^2-n^2)(2\rho^2+1)(1+x)+(m^2+n^2)(1+x)\right] (1-x)
d\phi^2  \nonumber \\ &+ \frac{1}{2V}  \left[
  4\rho^2(\rho^2+1)+(m^2-n^2)(2\rho^2+1)(1-x)+(m^2+n^2)(1-x)\right] (1+x)
d\psi^2.  \nonumber
\end{align}
If we define
\begin{equation}\label{susyangles}
\phi = \frac{1}{2}(\varphi-\tau), \quad \psi = \frac{1}{2}(\varphi+\tau),
\end{equation}
we can also rewrite \eqref{xybase} in a fibered form,
\begin{equation}
\bar ds_4^2 = \frac{[4\rho^2 (\rho^2+1) + (m-n)^2 (1-x^2)]}{4V} (d\tau +
\delta d \varphi)^2 + ds_3^2, 
\end{equation}
where 
\begin{equation} \label{3ddelta}
\delta = \frac{ [4\rho^2 (\rho^2+1) x +
  (m^2-n^2)(1-x^2)(2\rho^2+1)]}{[4\rho^2(\rho^2+1)+(m-n)^2 (1-x^2)]}
\end{equation}
and
\begin{equation} \label{3dbase}
ds_3^2 = \frac{d\rho^2}{\rho^2+1} + \frac{dx^2}{4(1-x^2)} + \frac{(1-x^2)
  \rho^2(\rho^2+1)}{[4\rho^2(\rho^2+1) +
  (m-n)^2 (1-x^2)]} d\varphi^2. 
\end{equation}
This fibered form is useful for comparison to the base metric in the
supersymmetric case, which is conventionally written in a similar
fibered form, and also for analysing the dimensional reduction to four
dimensions. The coordinates $\phi$ and $\psi$ are $2\pi$ periodic.
This implies that $\tau$ and $\varphi$ have periodicities
\begin{equation} \label{tpper}
(\tau, \varphi) \sim (\tau- 2\pi, \varphi + 2\pi), \quad \tau \sim
\tau + 4\pi. 
\end{equation}
The fermions will be antiperiodic under $(\tau, \varphi) \sim (\tau-
2\pi, \varphi + 2\pi)$ and periodic under $\tau \sim \tau + 4\pi$. 
The functions $\tilde H_i$ appearing in the metric become
\begin{equation}
\tilde H_i = 2c [2\rho^2 +1 - (m^2-n^2) x +E_i],
\end{equation}
where we introduce the convenient constants
\begin{equation}
E_i \equiv \sqrt{ \frac{Q_i^2}{4c^2} + [(m^2-n^2)^2 - 2(m^2+n^2)
  +1]}. 
\end{equation}
It was shown in~\cite{rossnonsusy} that $\tilde H_i>0$ everywhere which implies $E_i > (m^2-n^2)-1$. The
gauge fields are
\begin{align}
A_i = &\tilde H_i^{-1} \left\{ Q_i dt - \sqrt{\frac{Q_j
      Q_k}{Q_i}} \frac{c}{\sqrt{mn}} 
    \left[ n(m^2-n^2+1+E_i)(1-x) d\phi \right. \right. \\
\nonumber &+ \left. \left. m(m^2-n^2-1-E_i)(1+x) d\psi
    \right] \right\}. 
\end{align}

We can also write the ADM mass and angular momenta of the
five-dimensional asymptotically flat solution in the Einstein frame
(in units where $4G^{(5)}/\pi=1$) in terms of the integer parameters
as
\begin{equation}
M_{ADM} = 2c (E_1 + E_2 + E_3), 
\end{equation}
\begin{equation}
J_\psi = -\sqrt{\frac{m}{n}} \sqrt{Q_1 Q_2 Q_3}, \quad J_{\phi} =
\sqrt{\frac{n}{m}} \sqrt{Q_1 Q_2 Q_3}.
\end{equation}

This coordinate transformation provides us with a nice form of the
metric, which makes the special nature of the soliton solutions
evident. We see that the base metric is completely independent of the
charges or any length scale. The full metric is written in terms of
$m,n$, the length scale $c$, and the charges $Q_i$, which enter
explicitly only through the functions $\tilde H_i$. However, $c$ is
not an independent length scale: it should be determined in terms of
the charges and $m,n$. In the supersymmetric case $m=n+1$, we
had~\cite{mathur3} 
\begin{equation} \label{csusy}
c_{susy} = \frac{1}{4 n(n+1)} \frac{Q_1 Q_2 Q_3}{Q_1 Q_2 + Q_1 Q_3 +
  Q_2 Q_3}. 
\end{equation}
We should determine how $c$ is related to the charges in the
non-supersymmetric case. We first note that \eqref{cdef} is equivalent
to
\begin{equation}
c = \frac{M}{4 mn} (s^{-2} - s^{2})^{-1}.
\end{equation}
Since $s^2$ is determined in terms of $m,n$ by \eqref{smn}, the
problem is simply to determine $M$ in terms of the charges $Q_i$ and
$m,n$. Using
\begin{equation}
  Q_i = M \sinh \delta_i \cosh \delta_i = M \frac{\tanh
    \delta_i}{1 - \tanh^2 \delta_i}, 
\end{equation}
we can express $s^2 = \tanh \delta_1 \tanh \delta_2 \tanh \delta_3$ in
terms of the $Q_i$ and $M$. We want to solve this for $M$ as a
function of $s^2$ and the charges. The equation can be rearranged to
find that the combination $\bar M = M (s^{-2} -s^{2})^{-1} = 4mn c$
satisfies
\begin{align} 
  &(s^{-2}-s^2) Q_1 Q_2 Q_3 \bar{M}^3 (Q_1 \bar M + Q_2 Q_3)(Q_2 \bar
  M + Q_1 Q_3)(Q_3 \bar M + Q_1 Q_2) \nonumber \\ &+[2 Q_1^2 Q_2^2 Q_3^2 (Q_1^2 + Q_2^2 +
  Q_3^2) - (Q_1^4 Q_3^4 + Q_1^4 Q_2^4 + Q_2^4 Q_3^4)]
  \bar M^4 + 8 Q_1^3 Q_2^3 Q_3^3 \bar M^3 \nonumber \\ &+ 2 Q_1^2 Q_2^2 Q_3^2 (Q_1^2
  Q_2^2 + Q_1^2 Q_3^2 + Q_2^2 Q_3^2) \bar M^2 - Q_1^4 Q_2^4 Q_3^4 = 0.
  \label{ceq}
\end{align}
In the supersymmetric case, $s=1$, the first line is absent, so one
can see that~\eqref{csusy} satisfies this equation.\footnote{The
  equation becomes $[(Q_1 Q_2 + Q_1 Q_3 + Q_2 Q_3) \bar M - Q_1 Q_2
  Q_3] [ (Q_1 Q_2 - Q_1 Q_3 - Q_2 Q_3) \bar M - Q_1 Q_2 Q_3] [(Q_1 Q_2
  - Q_1 Q_3 + Q_2 Q_3) \bar M + Q_1 Q_2 Q_3)] [(Q_1 Q_2 + Q_1 Q_3 -
  Q_2 Q_3) \bar M + Q_1 Q_2 Q_3] =0.$} In the non-supersymmetric case,
there is no explicit solution for $c$ in general, but it is determined
implicitly in terms of the charges by~\eqref{ceq}. Explicit solutions
are possible in special cases: If all three charges are equal, we can
write
\begin{equation}
c = \frac{Q}{4mn} \frac{s^{-2/3} - s^{2/3}}{s^{-2} - s^2}.
\end{equation}
If one of the charges is much smaller than the other two, say $Q_3 \ll
Q_1, Q_2$, then 
\begin{equation}
c \approx \frac{Q_3}{4mn}. 
\end{equation}
This limit is interesting because it is the regime where the type IIB
solution has a near-core AdS$_3 \times S^3$ geometry.

\subsection{Alternative coordinates}
\label{alt}

The coordinates above provide the simplest description of the
solution, but they are not directly related to the coordinates used
for the supersymmetric solutions in~\cite{mathur3}. It is therefore
useful to introduce an alternative coordinate system which makes the
connection to the supersymmetric case clearer. These coordinates will
also be adapted to studying the local structure near one of the
orbifold singularities, as we will see later.

We therefore introduce new coordinates $r, \theta$ through
\begin{equation}
r = \frac{(\bar r^2- r_+^2)}{4} + c \sin^2 \bar \theta = c [ \rho^2 +
\frac{1}{2}(1-x)], 
\end{equation}
\begin{equation}
 r \cos^2 \frac{\theta}{2} = \frac{(\bar r^2 - r_+^2)}{4}
\cos^2 \bar \theta = \frac{c}{2} \rho^2 (1+x).
\end{equation}
The range of the coordinates is $r \geq 0$, $0 \leq \theta \leq \pi$.
One could similarly define coordinates centered on the singularity at
$\rho=0$, $x=-1$ by 
\begin{equation}
r_c = c [ \rho^2 + \frac{1}{2}(1+x)],
\end{equation}
\begin{equation}
 r_c \cos^2 \frac{\theta_c}{2} = \frac{c}{2} \rho^2 (1-x),
\end{equation}
but we will focus henceforth on the $r,\theta$ coordinates. It is
useful to retain $r_c$, however, thought of as a function of $r,
\theta$ given by
\begin{equation}
r_c^2  = r^2 + 2 r c \cos \theta + c^2;
\end{equation}
then the inverse coordinate transformation is 
\begin{equation} 
\label{invtrans}
2c \sin^2 \bar \theta = (r - r_c) + c, \quad \bar r^2 - r_+^2 =
2(r + r_c - c),
\end{equation}
or
\begin{equation}
\rho^2 = \frac{(r+r_c-c)}{2c}, \quad x = \frac{(r_c -r)}{c}. 
\end{equation}
These coordinates have some nice properties even for the general
metric \eqref{4dbasegen}:
\begin{equation}
g(\bar r) = 4 (r + r_c)^2 - 4c^2,
\end{equation}
which can be used to show
\begin{equation}
g(\bar r) \sin^2 \bar \theta \cos^2 \bar \theta = 4 r^2 \sin^2 \theta  
\end{equation}
and
\begin{equation}
\frac{\bar r^2 d\bar r^2}{g(\bar r)} + d\bar \theta^2 = \frac{1}{4 r r_c}
(dr^2 + r^2 d \theta^2),
\end{equation}
so this coordinate transformation casts this part of the metric in a
conformally flat form (note that we get a conformally flat form even
before restricting to the smooth soliton solutions). The base metric
used in the supersymmetric solutions differs from the one used up to
now by a conformal factor; define
\begin{equation}
ds_4^2 = 2c \sqrt{V} \bar{ds}_4^2.
\end{equation}

The eleven-dimensional metric is then
\begin{equation} \label{rtmetric}
ds^2_{11} = -(Z_1 Z_2 Z_3)^{-2/3} (dt + k)^2 + (Z_1 Z_2 Z_3)^{1/3}
ds_4^2 + ds^2_{T^6}, 
\end{equation}
with 
\begin{equation}
Z_i \equiv \frac{\tilde H_i}{2c \sqrt{V}} = \frac{1}{c\sqrt{V}}[ r+r_c
+ (m^2-n^2) (r-r_c) + c E_i],
\end{equation}
\begin{equation}
V= \frac{1}{c^2} 
[(m^2-n^2)(r - r_c) +  (r +r_c)]^2 - [
  (m^2-n^2)^2 - 2(m^2+n^2) + 1],
\end{equation}
and
\begin{align}
k = &\frac{\sqrt{Q_1 Q_2 Q_3}}{8c^3 \sqrt{mn} V}\left\{ -m [
  (m^2-n^2)(r - r_c +c) + (r + r_c -c)
  ](r_c - r + c) (d\tau + d\varphi) \right. \nonumber \\ &+
  \left. n
  [(m^2-n^2)(r-r_c-c) + (r+r_c-c)]
  (r - r_c + c) (-d\tau +
  d\varphi) \right\}.
\end{align}
The base metric here is (using the angular coordinates
\eqref{susyangles} appropriate for writing the base as a fibration)
\begin{equation} \label{4dbase}
{ds}_4^2 = H^{-1} \sigma^2 + H \left[ \gamma (dr^2 + 
  r^2 d\theta^2) + r^2 \sin^2\theta d\varphi^2
  \right], 
\end{equation}
with
\begin{equation} \label{hbase}
H = \frac{2c \sqrt{V}}{[ ((r + r_c)^2 - c^2) -
  (m-n)^2 ((r - r_c)^2 - c^2)]}, 
\end{equation}
\begin{equation}
\gamma = \frac{[ ((r + r_c)^2 - c^2) -
  (m-n)^2 ((r - r_c)^2 - c^2)]}{4 r r_c}, 
\end{equation}
and
\begin{equation}
\sigma = d\tau - \frac{[((r + r_c)^2 - c^2)
  (r - r_c) + ((r - r_c)^2 - c^2)
  (m^2-n^2) (r + r_c)]}{c [ ((r + r_c)^2 - c^2) -
  (m-n)^2 ((r - r_c)^2 - c^2)]} d\varphi.
\end{equation}

\subsection{Relation to the supersymmetric case}

The metric \eqref{rtmetric} is in the form used in writing the supersymmetric solutions, so we can make a detailed comparison to the supersymmetric case. If we set $m=n+1$, this metric should reduce
to the familiar two-center supersymmetric solution
of~\cite{mathura,mathurb,mathur3}. Indeed, for $m=n+1$, we will have $\gamma
=1$, $V = 4[(n+1) r - n r_c]^2$,  
\begin{equation}
H = \frac{n+1}{r_c} - \frac{n}{r}, 
\end{equation}
and
\begin{equation}
\sigma = d\tau +\left[(n+1)\cos \theta_c  - n
    \cos \theta \right] d\varphi,
\end{equation}
so the base space \eqref{4dbase} reduces to the usual Gibbons-Hawking
base of the two-center supersymmetric solution. This base space plays
an important role in the study of the supersymmetric solutions.

Compared to the supersymmetric cases, the obvious novelty in these
solutions is that the base metric \eqref{4dbase} is no longer of the
Gibbons-Hawking form. Indeed, although it is still a $U(1)$ fibration
over a three-dimensional base space, that space is no longer flat, or
even conformally flat, and the function $H$ does not appear to satisfy
a harmonic equation on this base. There is no clear sign of any
linear structure in the equations satisfied by this solution which
could be exploited to generate multi-center solitons as in the supersymmetric case. 

In the supersymmetric case, an important insight gained from the
analysis of the two-center solutions in~\cite{mathur3} was that the
signature of the base space switched where $H = 0$. That is, the base
space was not really a hyper-K\"ahler manifold, but only pseudo-hyper
K\"ahler. There is thus a much bigger space of possibilities for the
base metric. In the form \eqref{4dbase}, we still have that for $f
<0$, $H <0$, and for $f > M$, $H>0$, so the base has regions of
positive and negative signature. However, because $V <0$ in the
intermediate region $0 < f < M$, $H$ as defined in \eqref{hbase}, and
hence the 4d metric \eqref{4dbase}, will become imaginary in this
intermediate region. Just as in the supersymmetric case, the full
geometry \eqref{rtmetric} has real components and a definite signature
everywhere; this imaginary function is just an artifact of how we have
chosen to decompose the metric into a base and fiber. It hence does
not prevent us from using this decomposition to analyse the equations
of motion, analogous to what was done for the supersymmetric case. 

This suggests that if we wanted to look for further non-supersymmetric
solutions using this kind of decomposition, we would need to allow
possibilities which included a region where the base metric becomes
imaginary. However, the absence of any nice harmonic behaviour in this
form of the solution suggests that, unlike in the supersymmetric case,
this decomposition may not be particularly useful.

\section{Structure of the solution}
\label{str}

We now analyse the structure of the solitons in this M-theory form,
using the adapted coordinates. In~\cite{rossnonsusy}, it was shown
that the type IIB solution~\eqref{fibred} was completely smooth, but the
dualities we have performed mix up the gauge fields and the geometry,
so they can change the smoothness properties of the metric. We focus
on analysing the coordinate singularities at $\rho=0$ or $x = \pm 1$. We
will see that there are orbifold singularities in the M-theory form.
We will also see that the adapted coordinates are convenient for
understanding the structure of the solution.

It is clear from \eqref{xybase} that at $x =1$, the circle along
$\partial_\phi$ shrinks smoothly to zero size, while at $x=-1$, the
circle along $\partial_\psi$ shrinks smoothly to zero size. The
component of the gauge field along the circle which shrinks goes to
zero, so the gauge fields are also smooth there.

At $\rho=0$, a circle which is a combination of these two shrinks; to
analyse this region, it is convenient to introduce a different set of
angular coordinates. We define
\begin{equation}
\phi' = - l \phi - k \psi, \quad \psi' = n \phi + m \psi,
\end{equation}
where $k, l$ are integers such that $kn - ml =1$. Since we assume $m,
n$ are coprime, we can always find such integers.\footnote{If
$m ,n$ are not coprime, we could write $m = a \bar m$, $n = a
\bar n$, with $\bar m, \bar n$ coprime, and proceed as above, with
$k\bar n - l \bar m =1$. In this
case, there will be a $\mathbb{Z}_a$ orbifold singularity at $\rho=0$.} We
introduce them so that the identifications $\phi \sim \phi + 2\pi$,
$\psi \sim \psi + 2\pi$ are equivalent to $\phi' \sim \phi' + 2\pi$,
$\psi' \sim \psi' + 2\pi$.  In these coordinates, 
\begin{equation}
  k = \frac{\sqrt{Q_1 Q_2 Q_3}}{ \sqrt{mn} cV} \left[ mn \rho^2 d\phi' +
    \frac{1}{4} [ -(m^2-n^2)(1-x^2)-2\rho^2x + 2(kn+ml) \rho^2] d\psi'
  \right],
\end{equation}
and
\begin{align}
 d\bar{s}_4^2 &= \frac{d\rho^2}{(\rho^2+1)} + \frac{1}{4} \frac{dx^2}{(1-x^2)}
+ \rho^2 d\phi'^2 + \frac{\rho^4}{V}[-(m^2-n^2)x - (m^2+n^2) + 2(\rho^2+1)]
d\phi'^2 \\ &+ \frac{\rho^2}{V} [-(m^2-n^2)(km-nl)(1-x^2) -2(km-nl)x(\rho^2+1)+2
  (km+nl)(\rho^2+1)] d\phi' d\psi'  \nonumber \\ &+  \frac{(1-x^2)}{V}
  d\psi'^2 \nonumber  \\ &+
\frac{\rho^2}{V} [-(k^2-l^2)(m^2-n^2)(1-x^2) -2(k^2-l^2) x(\rho^2+1) + 2
  (k^2+l^2)(\rho^2+1)] d\psi'^2.  \nonumber  
\end{align}
Thus at $\rho=0$, the circle along $\partial_{\phi'}$ shrinks smoothly
to zero size.

The component of the gauge field along $\partial_{\phi'}$ is 
\begin{equation}
(A_i)_{\phi'} = - \sqrt{\frac{Q_j Q_k}{Q_i}} \sqrt{mn} \left( 1 -
  \frac{4c\rho^2}{\tilde H_i} \right).   
\end{equation}
As $\rho \to 0$, it goes to a constant, which gives a non-zero
holonomy around the shrinking circle, $\oint_{\phi'} A_i = - 2\pi
\sqrt{\frac{Q_j Q_k}{Q_i}} \sqrt{mn}$. Can this be removed by a large
gauge transformation? In the type IIB picture, the gauge field associated
with $q_p$ comes from a Kaluza-Klein reduction, so we know that $A_p
\to A_p + R d\phi'$ is an allowed large gauge transformation. Using
(\ref{mass}-\ref{period}), we can show $R = \sqrt{\frac{q_1 q_5}{q_p}}
\sqrt{mn}$. Regarding each of the gauge fields in turn as arising as
the Kaluza-Klein gauge field in a distinct reduction from type IIB, we can
argue that
\begin{equation}
A_i \to A_i + \sqrt{\frac{Q_j Q_k}{Q_i}} \sqrt{mn} d\phi'
\end{equation}
are indeed allowed large gauge transformations. That is, the gauge
group is a circle of size $2\pi \sqrt{\frac{Q_j Q_k}{Q_i}}
\sqrt{mn}$, and the holonomy around any cycle takes values in this
circle, so any holonomy that is an integer multiple of the size of the
circle is actually equivalent to zero holonomy. Thus, the gauge fields
have zero holonomy up to gauge transformations, and the gauge fields
are also smooth at $\rho=0$.

We should finally consider what happens at $\rho=0$, $x = \pm 1$,
where two circles shrink simultaneously. There will be orbifold
singularities at these points.

\subsection{Orbifold singularities}
\label{orbi}

The most interesting feature of the non-supersymmetric solitons in
this five-dimensional picture is the way in which they differ from the
supersymmetric case in their local structure near the points $\rho=0$,
$x=\pm 1$. As in the supersymmetric case, there are orbifold
singularities at these points. However, in the present case these are
non-supersymmetric orbifolds. Hence, the supersymmetry is broken not
just in the full asymptotically flat solution, but even by the local
solution describing the geometry near the singular points. This is
somewhat surprising, as supersymmetry is not broken locally in the type IIB
solution. In this section, we will explore the local structure,
showing that the supersymmetry is broken locally, and explain the
relation to the type IIB solution. We conclude from this local breaking of
the supersymmetry that the picture of the smooth supersymmetric
solutions as made up of 1/2 BPS `atoms' \cite{4dmicro} does not extend
to the present case.

Let us consider the point $\rho=0$, $x=1$, and study it in the $r,\theta$
coordinates introduced in section \ref{alt}, where it corresponds to
$r=0$. Near $r=0$, $V \approx 4 n^2$, $\tilde H_i \approx
2c[-(m^2-n^2)+1+E_i]$ is a constant, and $k \approx 0$, so the full
geometry decomposes as a flat $\mathbb R^{1,6}$ cross the
four-dimensional base space \eqref{4dbase}. In the base space, $r_c
\approx c + r \cos \theta$,
\begin{equation}
\gamma \approx  \frac{1}{2} [ (1+\cos \theta) + (m-n)^2 (1-\cos \theta)],
\end{equation}
and
\begin{equation}
H \gamma = \frac{2c \sqrt{V}}{4 r r_c} \approx \frac{n}{r}.
\end{equation}
Finally, the fiber is
\begin{equation}
\sigma \approx d\tau + \frac{1}{2 \gamma} [ (1+\cos \theta) +
  (m^2-n^2) (1-\cos\theta) ] d\varphi.
\end{equation}
We introduce a new radial coordinate $\tilde r = \sqrt{r}$, so the base
space can be written, up to an irrelevant constant scale factor, as
\begin{equation} \label{local4d}
ds_4^2 = d \tilde r^2 + \tilde r^2 (\frac{d\theta^2}{4} +
  \frac{1}{4\gamma} \sin^2 \theta d\varphi^2 + \frac{\gamma}{4n^2}
  \sigma^2).   
\end{equation}
The geometry looks locally like a cone over
\begin{equation} \label{lens}
d\Sigma^2 = \frac{d\theta^2}{4}+
  \frac{1}{4\gamma} \sin^2 \theta d\varphi^2 + \frac{\gamma}{4n^2}
  \sigma^2.   
\end{equation}
This is an orbifold of $S^3$. If we start from $S^3$ in the standard
form
\begin{equation} \label{s3lens}
d\Sigma^2 = \frac{d\theta^2}{4}+
  \frac{1}{4} \sin^2 \theta d\varphi'^2 + \frac{1}{4} (d \tau' - (1-
  \cos \theta) d\varphi')^2,
\end{equation}
we can obtain \eqref{lens} by the transformation 
\begin{eqnarray}
\tau' &=& \frac{1}{n}(\tau + \varphi), \\
\varphi' &=& -\frac{(m-n-1)}{2n} \tau - \frac{(m+n-1)}{2n} \varphi.
\end{eqnarray}
The periodicities \eqref{tpper} of $\tau$ and $\varphi$ in our
solution then imply that $\tau'$ and $\varphi'$ are identified under
$\varphi' \sim \varphi' - 2\pi$ and $(\tau',\varphi') \sim
(\tau'+4\pi/n, \varphi'-(m-n-1) 2\pi/n)$. This defines a freely-acting
$\mathbb{Z}_n$ quotient of $S^3$, which is referred to in the
mathematical literature as a lens space, denoted
$L(n,m)$~\cite{wiki:lens,hatcher02}.\footnote{Note that the lens
  spaces $L(n,m)$ and $L(n,m')$ are homeomorphic for $m = \pm m'$ mod
  $n$ or $mm' = \pm 1$ mod $n$.}  In the supersymmetric case $m =
n+1$, this is $L(n,1)$, and the second identification becomes $\tau'
\sim \tau' + 4\pi/n$, so the quotient just acts on the canonical fiber
of the $S^3$. This is no longer true in the non-supersymmetric cases.

There is thus a $\mathbb{Z}_n$ orbifold singularity at $r=0$. We can
make this manifest by defining complex coordinates 
\begin{equation}
u_1 = \tilde r \sin \frac{\theta}{2} e^{i \tau'/2} = \tilde r \sin
\frac{\theta}{2} e^{i (\tau + \varphi)/2n},
\end{equation}
\begin{equation}
u_2 = \tilde r \cos \frac{\theta}{2} e^{i (\tau' - 2\varphi')/2} =
\tilde r
\cos \frac{\theta}{2} e^{i [(m-n)\tau + (m+n)\varphi]/2n},
\end{equation}
in terms of which the local geometry is simply flat $\mathbb{C}^2$,
and the identification $\tau \sim \tau + 4\pi$ acts as
\begin{equation} \label{orbifold}
u_1 \sim u_1 e^{2\pi i/n}, \quad u_2 \sim u_2 e^{2\pi i (m-n)/n}. 
\end{equation}
The geometry is thus a non-supersymmetric $\mathbb{C}^2/\mathbb{Z}_n$
orbifold of the kind discussed for example in~\cite{aps,Harvey:2001wm}. The
supersymmetry is broken except in the special case $m=n+1$; the
identifications \eqref{orbifold} do not preserve any Killing spinors.
If $m=n+1$, it is the usual supersymmetric $\mathbb{C}^2/\mathbb{Z}_n$
orbifold (note the fermions are periodic under $\tau \sim \tau + 4\pi$
in the full geometry, so we have the supersymmetry-preserving choice
of spin structure on the orbifold).

However, this is not the end of the story; we also need to consider the
gauge fields. Near $r=0$,
\begin{equation}
A_i = \frac{Q_i}{\tilde H_i} dt + \sqrt{\frac{Q_j Q_k}{Q_i}}
 \frac{\sqrt{mn}}{2} d\tau' .  
\end{equation}
Thus, the holonomy of the gauge field around the orbifold circle is
\begin{equation} \label{orbihol}
\oint_{\tau'} A_i = 2\pi \sqrt{\frac{Q_j Q_k}{Q_i}}
 \sqrt{mn} \frac{1}{n}. 
\end{equation}
This is a fractional holonomy; it cannot be removed by a large gauge
transformation, which can only shift the holonomy by an integer
multiple of $2\pi \sqrt{\frac{Q_j Q_k}{Q_i}} \sqrt{mn}$, as argued in
the previous section.\footnote{Note that the gauge field is however
  well-behaved in the covering space $S^3$, as we would in general
  expect. That is, this gauge field can be thought of locally as
  arising from the orbifolding of a well-behaved gauge connection on
  the covering space.} 

The presence of this holonomy makes this orbifold qualitatively
different from the one considered in~\cite{aps}, even though the
geometry is the same. It makes the total space of the gauge bundle
regular at $r=0$. Although the direction the orbifold acts on in the
base shrinks to zero size at $r=0$, the orbifold also involves a shift
along the fiber given by \eqref{orbihol}, so there is a free action on
the total space. Thus, the local orbifold singularity will be
``frozen''; as in~\cite{deBoer:2001px}, the presence of this
non-trivial holonomy prevents us from resolving the singularity by
deforming it to a smooth ALE space with trivial fundamental group.
Note that the singularity is frozen even in the supersymmetric case
$m=n+1$. This freezing of the singularity also suggests that there is
no tachyon in the winding sectors in this non-supersymmetric orbifold 
as there is no natural endpoint for the condensation of such a
tachyon. Indeed, we will see below that there is no tachyon, by
considering the relation to the type IIB description.

The fact that the total space of the gauge bundle is smooth also
explains how this non-supersymmetric orbifold can arise from a smooth
geometry in six dimensions. In the Kaluza-Klein reduction to five
dimensions, the fiber always has finite size, so we might be surprised
that there is a singularity in the base, but the fractional holonomy
explains how this arises: even though the total space is smooth and
the fiber is not degenerating, the connection is not well-behaved at
this point.

Since the type IIB solution is smooth, it is also supersymmetric locally,
in a neighbourhood of any point,\footnote{And if we take one charge
small, the type IIB solution will have a near-core region which is
approximately AdS$_3 \times S^3$, so supersymmetry is actually
restored in the whole near-core region.} but the orbifold we
obtained locally at $r=0$ is non-supersymmetric. It is useful to
consider carefully what happens to the supersymmetry under the duality
from the type IIB to the M-theory picture. The key step is the reduction
to a five-dimensional solution from the type IIB solution \eqref{fibred} 
on $S^1_y \times T^4$. Recall that in the type IIB solution, there is a circle
which shrinks smoothly to zero size at $\bar r =   r_+$. The
solution thus has a unique spin structure, which is antiperiodic
around this circle, and also around each of the two contractible
$S^1$s in $S^3$ (parametrized by the coordinates $\phi$ and $\psi$
above). Imposing the latter two conditions, the Killing spinor must be
of the general form
\begin{equation}
  \epsilon = e^{i\epsilon_1  \alpha \frac{y}{2R}} e^{i\epsilon_2 
    \frac{\phi}{2} + i\epsilon_3  
    \frac{\psi}{2}} f(\bar r, \bar \theta) \epsilon_0.
\end{equation}
for some real $\alpha$ and constant spinor $\epsilon_0$, with three
independent sign choices $\epsilon_1, \epsilon_2, \epsilon_3= \pm 1$.
To impose the antiperiodicity on the degenerating circle, note that
going once around the circle which shrinks to zero at $\bar r=  r_+$ corresponds to going once around the $y$ circle, $-m$ times
around the $\phi$ circle, and $n$ times around the $\psi$ circle.
Thus, we must have $\alpha = 1 + \epsilon_1 \epsilon_2 m - \epsilon_1
\epsilon_3 n$ to produce the correct antiperiodic behaviour around
this circle. That is, the Killing spinor will be
\begin{equation} \label{ks}
  \epsilon = e^{i (\epsilon_1 + \epsilon_2  m- \epsilon_3 n)
    \frac{y}{2R}} e^{i\epsilon_2 
    \frac{\phi}{2} + i\epsilon_3  
    \frac{\psi}{2}} f(\bar r, \bar \theta) \epsilon_0= e^{i \epsilon_1
    \frac{y}{2R}} e^{i\epsilon_2  ( m\frac{y}{2R} +
    \frac{\phi}{2})} e^{-i\epsilon_3  ( n\frac{y}{2R} -
    \frac{\psi}{2})} f(\bar r, \bar \theta) \epsilon_0. 
\end{equation}
We get a five-dimensional solution by reducing on the $y$ circle.  In
the case $m=n+1$, where the whole solution is supersymmetric, the
Killing spinor \eqref{ks} with $\epsilon_2 = \epsilon_3 = -\epsilon_1$
is constant around the $y$ circle. For the non-supersymmetric
solutions, the Killing spinor \eqref{ks} will not be constant around
the $y$ circle for any choice of signs. As a result, it does not give
rise to a Killing spinor in the lower-dimensional theory. To see this,
note that we obtain the five-dimensional gravitino from the constant mode of the six-dimensional
gravitino.\footnote{This is true if we work in the sector where $m+n$
  is odd. If $m+n$ is even, the spin structure on the six-dimensional solution does
  not give rise to a spin structure on the five-dimensional solution, as we are
  Kaluza-Klein reducing on a circle with antiperiodic boundary
  conditions for the fermions.}  Thus, we cannot obtain a
supersymmetry in the five-dimensional solution from the supersymmetry in the six-dimensional
solution: the Killing spinor which provides the supersymmetry
parameter for the six-dimensional solution descends to a spinor field on the five-dimensional
solution which is charged under the Kaluza-Klein gauge field, so it
cannot be the parameter for a variation of the gravitino, which is not
charged under the Kaluza-Klein gauge field.

That is, although the local six-dimensional geometry has a
supersymmetry even for non-supersymmetric solitons, this supersymmetry
is not visible in the five-dimensional solution. This is an example of
the phenomenon of ``supersymmetry without supersymmetry''
\cite{susywosusy}. It may seem surprising that there is such a
connection between whether the local orbifold singularities in the
five-dimensional solution preserve supersymmetry and whether the full
type IIB solution does; but in both cases, the condition to preserve
supersymmetry is that the Killing spinor is constant around the $y$
circle.\footnote{In the full type IIB solution, this arises from the
  fact that in the asymptotically flat region, the only possible
  Killing spinors are constant around the $y$ circle.}

This relation to the type IIB solutions also implies that there is no
instability of the local geometry. That is, there will not be any
winding tachyon modes of the type studied in~\cite{aps}.  The absence
of tachyons is clear from the type IIB solution: we would not expect
any tachyon modes in the smooth soliton geometry, and in the cases
where there is a near-core AdS$_3 \times S^3$ geometry, this
supersymmetric solution is known to have no tachyons.  Intuitively, we
can understand the effect of the holonomy as requiring twisted sector
strings to ``stretch'' along the fiber direction, making an additional
contribution to their energy and lifting the tachyons up to positive
mass-squared.

To summarize our analysis of the M-theory form of the solitons, we
have found that the five-dimensional solution is smooth except at two
points, where we have non-supersymmetric orbifold singularities. Thus,
the solution has a ``two-center'' structure, which is similar to the
simplest supersymmetric solitons. However, these centers do not
preserve any supersymmetry, even locally. This is one of the main
lessons from our analysis: the picture of the supersymmetric solutions
in \cite{4dmicro}, in which they were understood as built up of 1/2
BPS ``atoms'', {\it does not} extend to the non-supersymmetric
solutions of \cite{rossnonsusy}. Even the atoms are not
supersymmetric. It would be very interesting to have some further
characterization of what atoms may be possible.

\section{Reduction to four dimensions}
\label{4d}

In this section we examine Kaluza-Klein reductions of our solution to four
dimensions. Our ambition in considering these reductions was to find a
description of the non-BPS atoms in terms of brane systems in
IIA. Unfortunately, we have failed to achieve this, but it is perhaps
instructive to explain how it goes wrong.

We would like to have a reduction which is smooth away from the
centers at $\rho = 0$, $x= \pm 1$. Considering the local geometry near
$r=0$, this amounts to asking for a representation of the lens space
\eqref{lens} as an $S^1$ bundle over some smooth two-dimensional
manifold. In the supersymmetric case $m=n+1$, $\gamma =1$, and
\eqref{lens} is already of the desired form: the $S^1_\tau$ is fibered
over $S^2$. However, in the general case, there is a problem. Since
$\gamma(\theta = 0) = 1$, the two-dimensional space is still smooth at
$\theta=0$, but $\gamma(\theta=\pi) = (m-n)^2$, so the two-dimensional
space has a $\mathbb{Z}_{|m-n|}$ orbifold singularity along $\theta = \pi$. In
the full four-dimensional geometry, this will form a line conical
singularity connecting the two centers. 

Is there some other, inequivalent way to represent the lens space as
an $S^1$ bundle that avoids this problem? Start from the metric in the
manifestly locally $S^3$ form \eqref{s3lens}, and consider the general
$S^1$, which is
\begin{equation} \label{gens1}
(\tau', \varphi') \sim (\tau' + 4\pi \frac{e}{n}, \varphi' - 2\pi
\frac{(m-1) e}{n} + 2\pi f)
\end{equation}
for some coprime integers $e, f$ .\footnote{If $e, f$ are not
  coprime, the cycle is not primitive---it is an integer
  multiple of some other cycle, and we should consider
  instead the corresponding primitive cycle.} We introduce coordinates
$\bar \tau$ and $\bar \varphi$ through
\begin{equation} 
\tau' = \frac{e}{n} \bar \tau + 2 \frac{g}{n}  \bar \varphi, 
\end{equation}
\begin{equation}
\varphi' = \frac{1}{2} \left( -\frac{(m-1)e}{n} + f \right) \bar \tau
+ \left( -\frac{(m-1)g}{n} + h \right) \bar \varphi, 
\end{equation}
where $g,h$ are integers such that $eh-fg =1$. The $S^1$ cycle
\eqref{gens1} is along $\partial_{\bar\tau}$ in these coordinates,
and the identifications are $\bar \tau \sim \bar \tau + 4\pi$, $\bar
\varphi \sim \bar \varphi + 2\pi$. Rewriting \eqref{s3lens} in these
coordinates, it becomes
\begin{equation}
d \Sigma^2 = \frac{d\theta^2}{4}+
  \frac{1}{4G} \sin^2 \theta d\bar \varphi^2 + \frac{G}{4n^2}
  \bar \sigma^2, 
\end{equation}
where $\bar \sigma = d\bar \tau + F d\bar \varphi$ (the precise form of $F$ will not be necessary for our arguments), and 
\begin{equation}
G = \sin^2 \theta ((m-1) e -n f)^2 + [e + \frac{1}{2} (1-\cos \theta)
((m-1)e -nf)]^2. 
\end{equation}
Thus, we can make the two-dimensional metric smooth at $\theta=0$ by
choosing $e= \pm 1$, and we can make it smooth at $\theta = \pi$ by
choosing $me -nf = \pm 1$.\footnote{This corresponds to a reduction of
  the full solution along the $\psi'$ coordinate. We can easily see
  that the base metric~\eqref{xybase} will be smooth at $\rho=0$ for
  such a reduction, but will have conical singularities along $x = \pm
  1$.  Locally near $\rho=0$, these two reductions are analogous to
  the reductions of a flat metric $ds^2 = dr^2 + r^2 d\phi^2 + R^2
  (d\tau + d\phi)^2$ with the identifications $\phi \sim \phi +\pi$,
  $\tau \sim \tau + 2\pi$.  If we reduce this along $\tau$, we get a
  flat space with a $\mathbb{Z}_2$ orbifold. If we reduce instead
  along $\phi$, we get a smooth fluxbrane solution in two dimensions.}
But we cannot satisfy both conditions simultaneously unless $m = \pm
1$ mod $n$, that is, unless we are considering the lens space $L(n,1)$
which arises in the supersymmetric case.

The conclusion is thus that in the general case, the best we can do is
to make the solution smooth on one of the two axes near $r=0$. It
seems sensible to keep the solution smooth at large distances; we thus turn to considering in a little more detail the reduction along
$\partial_\tau$, which has a $\mathbb{Z}_{|m-n|}$ conical singularity
connecting the two centers. To do so, we first need to rewrite the
metric in the usual Kaluza-Klein form. The metric is of the general form
\begin{equation}
ds_5^2 = - A (dt + \beta (d\tau + \delta d\varphi) + \omega
d\varphi)^2 + B (d \tau + \delta d\varphi)^2 + C ds_3^2,  
\end{equation}
where $\delta$ was given in \eqref{3ddelta} and $ds_3^2$ was given in
\eqref{3dbase}.  It can be written as
\begin{equation}
ds_5^2 = (B-A \beta^2) \left[  d\tau + \delta d\varphi - \frac{A\beta}{B-A
    \beta^2} ( dt + \omega d\varphi) \right]^2
  - \frac{AB}{B-A\beta^2}(dt + \omega d\varphi)^2 +
  C ds_3^2. 
\end{equation}
If we then reduce along $\tau$, the last two terms will give the
four-dimensional geometry.

Thus, the three-dimensional metric in \eqref{3dbase} is the base space
for the four-dimensional geometry: we repeat it here,
\begin{equation}
ds_3^2 = \frac{d\rho^2}{\rho^2+1} + \frac{dx^2}{4(1-x^2)} + \frac{(1-x^2)
  \rho^2(\rho^2+1)}{[4\rho^2(\rho^2+1) +
  (m-n)^2 (1-x^2)]} d\varphi^2. 
\end{equation}
In the supersymmetric case $m=n+1$, the base metric~\eqref{3dbase} is
flat. In the non-supersymmetric case, it is not even conformally flat.
We can see directly that it is smooth along the axes which extend to
infinity, at $x= \pm 1$, and has a conical singularity between the two
centers, along the line at $\rho=0$.  Also, there are curvature
singularities in this base metric at $\rho=0$, $x= \pm 1$. These are
in addition to the curvature singularities in the full four-dimensional
metric which will arise from the fact that the dilaton is diverging
at these points.

To simplify the Kaluza-Klein reduction, it is better to change to the
coordinate $\tau' = \tau + \varphi$, so that the identifications are
simply $\varphi \sim \varphi + 2\pi$ and $\tau' \sim \tau' + 4\pi$.
From the Kaluza-Klein point of view, this is a gauge transformation
which shifts $\delta \to \delta\,'=\delta -1$. 

The Kaluza-Klein gauge field coming from this reduction is
\begin{equation}
A_{KK} = \delta\,' d\varphi - \frac{A\beta}{B-A
    \beta^2} ( dt + \omega d\varphi),
\end{equation}
where 
\begin{equation}
\omega = - \frac{\sqrt{Q_1 Q_2 Q_3}}{4c \sqrt{mn}} \frac{(m-n) \rho^2
(1-x^2)}{[4\rho^2(\rho^2+1)+(m-n)^2 (1-x^2)]},
\end{equation}
and after the above gauge transformation,
\begin{equation} \label{moddelta}
\delta\, ' =  \frac{ [4\rho^2 (\rho^2+1) (x-1) +
  (m^2-n^2)(1-x^2)2\rho^2 +2n(m-n)(1-x^2)]}{[4\rho^2(\rho^2+1)+(m-n)^2 (1-x^2)]}.
\end{equation}
Since $\omega =0$ at $\rho=0$ and at $x = \pm 1$, the second factor in
$A_{KK}$ will not contribute to holonomies at these points. We
therefore do not write it more explicitly.  The factor of
$B-A\beta^2$, which will give the dilaton of the four-dimensional
solution, is more complicated:
\begin{align}
B - A \beta^2 = &\frac{(\tilde H_1 \tilde H_2 \tilde H_3)^{1/3}}{4 
  V} [4\rho^2(\rho^2+1)+(m-n)^2 (1-x^2)] \\ &- \frac{Q_1 Q_2 Q_3}{(\tilde H_1
  \tilde H_2 \tilde H_3)^{2/3} 16  V mn} \{2\rho^2[(m+n)+x(m-n)] +
(m-n)^2(m+n) (1-x^2)\}^2. \nonumber
\end{align}
After a lot of algebra, and using the constraint \eqref{ceq} that
determines $c$ in terms of the $Q_i$, the factor of $V$ cancels, and
this can be written as
\begin{align}
B - A \beta^2 = &\frac{1}{(\tilde H_1
  \tilde H_2 \tilde H_3)^{2/3} } \left\{ \frac{Q_1 Q_2 Q_3}{16 mn}
  [4\rho^2 +(m-n)^2(1-x^2)] \right.  \\ \nonumber &\left.+2c^3 [2\rho^2+1 - (m^2-n^2)
  x + E_1 + E_2 + E_3] [4\rho^2(\rho^2+1)+(m-n)^2 (1-x^2)]  \right\}. 
\end{align}
This clearly vanishes at $\rho=0, x=\pm 1$. We will thus have
singularities of the four-dimensional metric and dilaton at these
points. It will not vanish at any other point, as $ E_i >
(m^2-n^2)-1$, so both terms are positive away from $\rho=0, x = \pm
1$. 

In the supersymmetric case, we would view the singularities at
$\rho=0, x=\pm 1$ as D6-branes. Here, however, an attempt to define the
charge carried by these singularities is obstructed by non-trivial
holonomies around the conical singularity. The Kaluza-Klein gauge
field which arises in the reduction from five to four dimensions has a
non-trivial holonomy around the conical line at $\rho=0$,
\begin{equation}
\oint_\varphi A_{KK} |_{\rho=0} = 4\pi \frac{n}{(m-n)}.
\end{equation}
Note that for $m=n+1$, this is an integer multiple of $4\pi$, and
hence gauge-equivalent to zero. This fractional holonomy in the
general case implies that there is a delta-function singularity in the
field strength $F_{KK}$ along $\rho=0$, and we cannot associate
separate Kaluza-Klein charges with the two singularities at $\rho=0$,
$x= \pm 1$. Thus, we cannot interpret the singularities as due to the
presence of D-branes at these points, since we cannot unambiguously
define brane charges associated with the singularities.\footnote{Note that the
total charge carried by the two singularities is still well-defined.
The integral of the flux over a sphere enclosing the whole line at
$\rho=0$ is $\int F_{KK} = \oint_\varphi A_{KK} |_{x=1} - \oint_\varphi A_{KK}
|_{x=-1} = 4\pi$,
so the structure carries one unit of KK monopole charge through a
surface at large distance. That is, the solution is asymptotically
flat in five dimensions.} 

Similarly, for the gauge fields $A_i$, we can decompose
\begin{equation}
A_i = A_i' + \alpha \left[  d\tau + \delta d\varphi - \frac{A\beta}{B-A
    \beta^2} ( dt + \omega d\varphi) \right], 
\end{equation}
and interpret $A_i'$ as the four-dimensional gauge field.\footnote{Note that
$A_i'$ is invariant under the gauge transformation made on $A_{KK}$,
so it does not matter if we use $\tau$ and the $\delta$ of
\eqref{3ddelta}, or $\tau'$ and the $\delta'$ from \eqref{moddelta} in
carrying out this calculation. We will use the former.} We find that
$A'_{i \varphi} = 0$ at $x = \pm 1$. Once again, $\omega$ vanishes
along $\rho^2=0$, so near the conical line singularity,
\begin{equation}
A'_{i \varphi} \approx A_{i \varphi} - \delta A_{i
  \tau} . 
\end{equation}
Hence at $\rho=0$ we find 
\begin{equation}
\oint_\varphi A_i' |_{\rho=0} = - 2\pi \sqrt{\frac{Q_j Q_k}{Q_i}} \sqrt{mn}
\frac{1}{m-n},
\end{equation}
showing the same fractional holonomy for these gauge fields as
well. As for the Kaluza-Klein gauge field, this implies that the
charges associated with the two singularities at $\rho=0$, $x = \pm 1$
are ambiguous.

We saw in the previous section that the M-theory solution has
non-supersymmetric orbifold singularities at $\rho = 0$, $x = \pm 1$.
We had hoped that the reduction to four dimensions would clarify the
interpretation of these singularities. However, the four-dimensional
solution has some new features (compared to the supersymmetric case)
which make it difficult to extract a brane interpretation of these
singularities. The four-dimensional solution has curvature
singularities only at $\rho = 0$, $x = \pm 1$; however, it also has a
line conical singularity connecting these two curvature singularities,
along $\rho = 0$. This conical singularity makes it impossible to
unambiguously assign charges to the two curvature singularities. Also,
the curvature singularities at $\rho = 0$, $x = \pm 1$ arise from a
divergence of the dilaton (as in the supersymmetric case) but also
from a singularity in the three-dimensional base metric at these
points. It is not clear what interpretation we could give to this
additional singularity from a brane construction point of view. The
geometry describing a D6-brane has a flat base space, and just gets a
curvature singularity from the divergence of the dilaton at the
brane's position. Thus, in the non-supersymmetric case, the centers do
not have a clear D-brane interpretation. It would be interesting to
see whether and how this local structure is changed for
non-supersymmetric solutions which are asymptotically flat in four
dimensions. 

\section*{Acknowledgements}
We thank Henriette Elvang, Ori Ganor, and Ashish Saxena for
discussions. We would also like to thank the Aspen Center for Physics
where part of this work was conducted.  The work of SFR was supported
in part by the EPSRC and PPARC.  The work of EGG was supported by the US
Department of Energy under contracts DE-AC03-76SF00098 and DE-FG03-91ER-40676
and by the National Science Foundation under grant PHY-00-98840.

\bibliographystyle{utphys}

\bibliography{DWsimon}

\end{document}